\documentclass{aa_no_copyright}  
\usepackage{graphicx}
\usepackage{txfonts}
\begin{document}
   \title{The new nebula in LDN\,1415 -- A cry from the cradle of a low-luminosity source\thanks{Based on observations performed at the Th\"uringer Landessternwarte Tautenburg.}}
   \author{B. Stecklum\inst{1}, S. Y. Melnikov\inst{1,2}, and H. Meusinger\inst{1} }
   \authorrunning{Stecklum et al.}
   \offprints{B. Stecklum}
   \institute{Th\"uringer Landessternwarte Tautenburg,
   		Sternwarte 5, D-07778 Tautenburg
	\and Ulugh Beg Astronomical Institute, Academy of Science of Uzbekistan, Astronomical str. 33, Tashkent 700051, Uzbekistan\\
              \email{stecklum@tls-tautenburg.de}}

   \date{Received 2006 Apr. 13; accepted 2006 Nov. 8}

   \abstract{}{A survey for candidate Herbig-Haro objects was performed to search for evidence of star formation in Galactic dark clouds}{For this aim true colour images were created from blue, red, and infrared DSS2 plates and inspected. Follow-up $I$-band, H$\alpha$,{and [S{\sc\,ii}]} CCD imaging as well as long-slit spectroscopy using the Tautenburg 2-m telescope was {carried out} \rm to verify candidate objects.}{In the case of LDN\,1415, the presence of a Herbig-Haro flow could be revealed {which is henceforth named HH\,892}\rm. In addition, an arcuate nebula was found which is barely seen on the DSS2 infrared plate (epoch 1996) and not detected in archival Kiso Schmidt data (epoch 2001). Thus, this nebula must have brightened by about 3.8\,mag in recent years.}{The nebula is associated with IRAS\,04376+5413. {The 2MASS images show a red counterpart of the IRAS source, designated as L1415-IRS}. Its morphology resembles that of a bipolar young object. The luminosity of this source integrated from 0.9\,$\mu$m to 60\,$\mu$m {in the low state\rm} amounts to {0.13\rm}\,$L_{\sun}$ for an assumed distance of 170\,pc. Thus it seems to be a young very-low mass star or it might even be of substellar mass. {The current brightness increase of the nebula is caused by a FUor- or EXor-like outburst as indicated by the presence of a P Cygni profile of the H$\alpha$ line}.  {L1415-IRS} is by far the least luminous member of the sparse sample of FUor and EXor objects.
} 
% 5 {} token are mandatory


   \maketitle
%
%________________________________________________________________

\section{Introduction}
{Herbig-Haro objects (HHOs) are a distinct tracer of star formation which allow to identify young stellar objects (YSOs) while they are still deeply embedded (\cite{ReiBal01})}. A survey for candidate HHOs associated with dark clouds was performed based on DSS2 plates utilising the fact that the filter transmission of the $R$ filter almost peaks at H$\alpha$ and the [S{\sc\,ii}]\,$\lambda\lambda$\,6717,\,6731 lines. Thus, in a true colour image based on blue, red, and infrared DSS2 plates, candidate HHOs show up since their colours, dominated by the emission lines, are very much different from those of stars. While the general results of this survey will be presented elsewhere, attention is being paid here to the particularly interesting case of LDN\,1415 (\cite{Lynds62}), an hitherto non-studied dark cloud. The cloud of opacity class 3 covers about 200$^{\square}$\arcmin{} {and} consists of three subclouds. {The only IRAS point source associated with LDN\,1415 is IRAS\,04376+5413, situated in the southern subcloud}. The identification of a candidate HHO in the {DSS2} survey led to follow-up CCD imaging and spectroscopy. The results of these observations, together with findings based on archival data, is the subject of the present paper. {A brief summary of the results was given by \cite{Steck06}}.

%__________________________________________________________________

\section{Data acquisition}
\subsection{Imaging}
Blue, red, and infrared images were retrieved from the STScI DSS2 server and combined to a true-colour image. Archival images taken with the KISO Schmidt telescope on 2001 January 22th were secured from the SMOKA archive (\cite{Baba02}). $J$-, $H$-, and {$K$s}-band images from the Two Micron All Sky Survey (2MASS, \cite{Skrut06}) were used for comparison with the optical images. The SCANPI tool was used to extract the mid- and far-infrared fluxes of the IRAS source. For this purpose, only IRAS scans within {a cross-scan distance of} less than 0\farcm5 from the position of the infrared source found on the 2MASS images were chosen. According to the SCANPI documentation, the peak flux of the fitted PSF template is considered to be representative {in the case of weak sources}.

CCD images of LDN\,1415 were taken on 2006 April 2 and 5th with the 2k$\times$2k prime focus camera at the 2-m telescope of the Th\"uringer Landessternwarte Tautenburg (TLS, diameter of the Schmidt {corrector} plate 1.34\,m). The instrument offers a field of view of 42\arcmin$\times$42\arcmin{} at the pixel scale of 1\farcs235. On the first night, $I$, H$\alpha$, {and [S{\sc\,ii}]} filters were used. {Because of deteriorating weather conditions only the first [S{\sc\,ii}] image turned out to be useful. The integration times amount to 180\,s for the broad-band and 1200\,s for the narrow-band exposures. The FWHM of the stellar images on the final H$\alpha$ frame amounts to 3\farcs2$\times$2\farcs8 (position angle of major axis 71\degr)}. On the second night, deeper images were obtained using $V$, $I$, and $z$ filters with an integration time of 900\,s each. Generally, two exposures were acquired per filter to allow for the reliable removal of cosmic ray events. {Further broad-band imaging was obtained on 2006 April 23. The standard stars SA104-338 and SA101-415 (\cite{Land92}) were observed for photometric calibration. Based on their measurements secondary standards were established within 10\arcmin of the position of IRAS\,04376+5413, using an inner aperture radius of 6\farcs2 and a sky annulus of [1.5, 2] times this size.} {In order to monitor the brightness of the source,  $R-$ and $I-$ band imaging was performed on 2006 Aug. 29, Sep. 29, and Oct. 16.} The images were flat-fielded with dome flats and astrometrically calibrated using the USNO-A2.0 catalogue (\cite{Monet98}).
%In order to estimate the distance of LDN\,1415, the molecular line data from the
%CO survey of \cite{Dame01} were used.

{
\subsection{Long-slit spectroscopy}
Long-slit spectroscopy of the HH flow and the nebula were obtained on 2006 Sep. 21 and Sep. 25  using the Nasmyth spectrograph at the 2-m telescope of the TLS equipped with a 2800$\times$800 pixel SITe CCD. Slit widths of 2\arcsec{} and 1\arcsec{} were applied which, together with the V100 grism, led to a resolution of R$\sim$2100 for the final spectra. The total integration time amounts to 8.4\,ks.  The wavelength calibration is based on night-sky lines. For this purpose, a template spectrum was generated by convolving the Osterbrock Sky Spectrum\footnote{Preparation of the Osterbrock Sky Spectrum files was supported by grant No. ATM-9714636 from the NSF CEDAR program} (www.nvao.org/NVAO/download/Osterbrock.html) to the observed resolution. The dispersion curve was derived by stretching the wavelength scale using a second order polynomial to make the template spectrum match the observed one in the least-squares sense for the wavelength region from 6200\,\AA\ to 6950\,\AA{}. The zero point of this calibration was checked using the sky spectrum as input and tied to the [O{\sc\,i}]\,6533\,\AA\ line which is a strong unblended sky line close to H$\alpha$. The accuracy of the radial velocity is in the order of 10\,km\,s$^{-1}$. The slit position angle (PA) of the Nasmyth spectrograph depends on declination. By chance, the declination of the target led to a slit PA which allowed to observe the HH flow and the nebula simultaneously.
}
%______________________________________________________________
\section{Results}

\subsection{The Herbig-Haro flow}
\begin{figure}[t]
%   \sidecaption
%        \includegraphics[width=12cm]{hals2fig.eps}
        \resizebox{\hsize}{!}{\includegraphics[width=12cm]{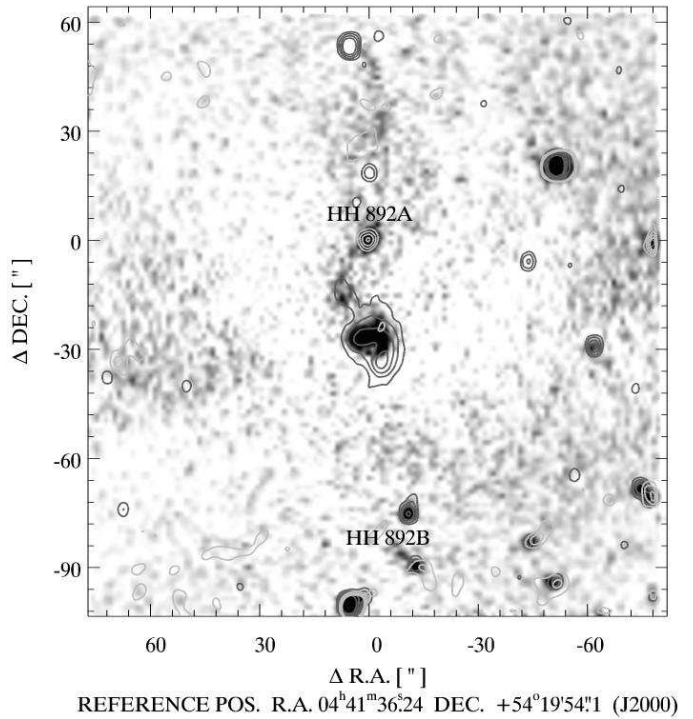}}
        \caption[]{High-contrast display of the continuum-subtracted H$\alpha$ image.{The light contours delineate continuum-subtracted [S{\sc\,ii}] emissionF while the dark contours represent the $I$-band image}. The HHOs are labeled. HH\,892A is located at the reference position [0\arcsec, 0\arcsec]. HH\,892B is situated at [$-$8\arcsec, $-$87\arcsec]. The northern part of the new nebula at [0\arcsec, $-$30\arcsec] is visible in H$\alpha$ too. }
        \label{Fig1}
   \end{figure}

On the DSS2 true-colour image of LDN\,1415, a HHO candidate close to IRAS\,04376+5413 was identified. Subsequent H$\alpha$ {and [S{\sc\,ii}]} imaging confirmed the presence of the emission line object which has no counterpart in the broad-band CCD images nor in 2MASS. The HHO was assigned No. {892} in the catalogue of Reipurth (\cite{Reip99}). Fig.\,\ref{Fig1} shows the continuum-subtracted H$\alpha$ image, based on an intensity scaling of the deep $I$-band image that yielded a good cancellation of the neighbouring red stars.  The remaining stellar images are those of less reddened foreground stars. The HHO is marginally resolved, {with an FWHM of 4\farcs4$\times$3\farcs2 (position angle 53\degr)}, and associated with faint wiggly {structure} to the north. {The less sensitive [S{\sc\,ii}] image shows the HHO at the 5$\sigma$ level.} About 1\farcm5 south of the HHO, a comma-shaped emission feature can be seen which is too faint to be visible on the {[S{\sc\,ii}] image and} the red DSS2 plate. However, it is well aligned with the components of the NIR object (see below) and the bright HHO. Thus it seems to be physically related to the nebula, {and is most likely the southern component of a HH flow}. {Henceforth the northern HHO will be called HH\,892A and the southern HH\,892B}.

The red DSS2 plate on which the HH\,892A was spotted first was taken on 1990 October 26th. {The comparison of the centroid positions of both images yields an upper limit for its annual proper motion of 0\farcs032 which corresponds to a tangential velocity of 25\,km/s for the likely distance of 170\,pc (see Sect.\,4)}. HH\,892A has the entry 1443$-$0136438 in the USNO-B1.0 catalogue \cite{Monet03} with $R_{\rm POSS-I}=19.59$\,mag, $R_{\rm POSS-II}=19.20$\,mag, and zero proper motion. {The CCD $R$ magnitude amounts to $19.65\pm0.43$\,mag.}

{
The spectrum of HH\,892A is shown in Fig.\,\ref{SpecFig}. It resembles that of a high-excitation HHO. For the fainter HH\,892B only the H$\alpha$ line was detected. Within the errors the radial velocities of both HHOs agree and are close to a $v_{LSR}$ of zero.

   \begin{figure}[]
        \resizebox{\hsize}{!}{\includegraphics[]{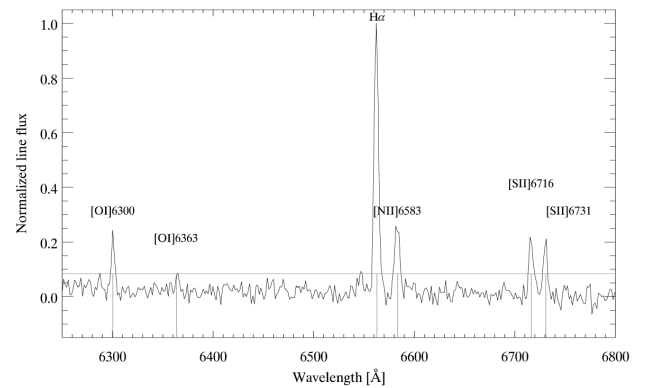}}
        \caption[]{Spectrum of HH\,892A. The identified lines are labeled. The horizontal line marks the 3$\sigma$ detection limit.}
        \label{SpecFig}
   \end{figure}

}

\subsection{A new nebula in LDN\,1415}
   \begin{figure}[t]
        \resizebox{\hsize}{!}{\includegraphics[]{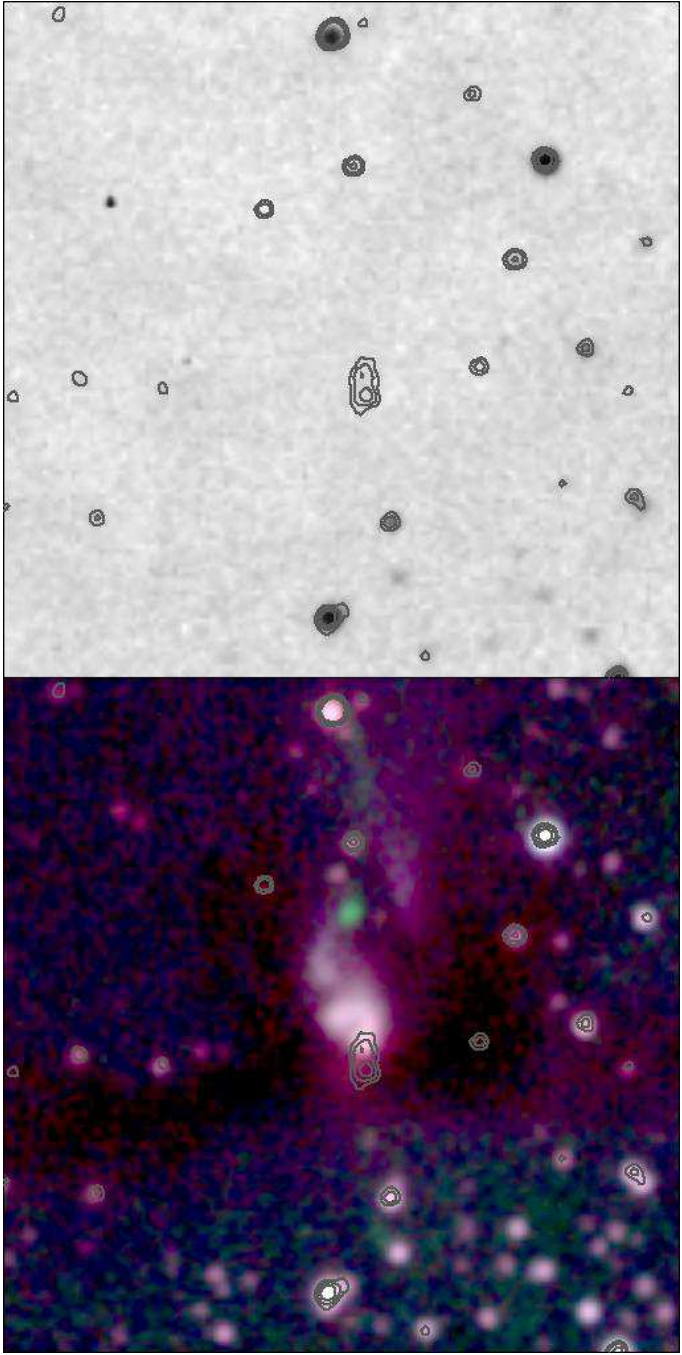}}
%        \resizebox{\hsize}{!}{\includegraphics[]{Kiso_TLS_2_grey.eps}}
        \caption[]{{Kiso $I$-band (top) and TLS RGB (bottom) images of the central region of LDN\,1415 (FOV 3\arcmin$\times$3\arcmin). The contours represent the 2MASS $Ks$ image at [4, 8, 16]\,$\sigma$}.  The TLS image is based on the stacked $I$, H$\alpha$, and $R$ frames. The new arcuate nebula is accompanied by extended faint emission to the north. The printed version of the paper shows the summed-up images.}
        \label{Fig2}
   \end{figure}  

 {The TLS CCD images show an arcuate nebula about 30\arcsec{} south of HH\,892A (Fig.\,\ref{Fig2} bottom).} This nebula is not visible down to the limiting magnitude of the DSS2 blue and red images. Inspection of the corresponding DSS1 image does not show any evidence of the nebula either. Thus the nebula seems to be the result of a recent process within LDN\,1415.  {Fig.\,\ref{Fig2} (top) illustrates that the Kiso I-band image taken in January 2001 also does not show any trace of this feature}.

{Based on the secondary photometric standards, total magnitudes of $R=15.9\pm 0.1$ and $I=14.6\pm 0.1$ were derived for the emission within an aperture of 10\arcsec{} radius centered on the nebula. The magnitudes of the brightness peak are $R=16.8\pm 0.1$ and $I=15.3\pm0.1$, respectively. Within the epoch difference ($\sim$250 days) of the TLS CCD imaging, no significant change in brightness was obvious. 
} 
On the POSS-II N infrared image which was taken as the last one of three POSS-II frames on 1996 December 12th, a patch of very faint emission is visible just south of the location of the current optical brightness peak. Although the emission is at the 2$\sigma$ level only, its spatial extent and the coincidence with the optically reddest part of the currently seen nebula indicate its a-posteriori detection. The total magnitude of {$I=18.4\pm 0.2$\,mag} was derived for the same aperture as for the CCD image. Thus, the brightness increase in the $I$-band amounts {to 3.77$\pm 0.25$\,mag. The $I$-band total magnitude in the low state corresponds to the limiting magnitude of the Kiso image. Therefore it can be concluded that the brightening occured within the last 5 years.
}

{
The optical spectrum of the nebula shows a red continuum. No other lines except H$\alpha$ were detected which is indicative of strong veiling. The latter displays a pronounced P Cygni profile displayed in Fig.\,\ref{FigSpecNeb}. The emission peak is redshifted at 70\,km\,s$^{-1}$ while the maximum absorption occurs at $-$210\,km\,s$^{-1}$. In particular there is no sign of the [S{\sc\,ii}]$\lambda$,$\lambda$6717,6731 lines which are tracers for HHOs knot and jets.

   \begin{figure}[h]
        \resizebox{\hsize}{!}{\includegraphics[]{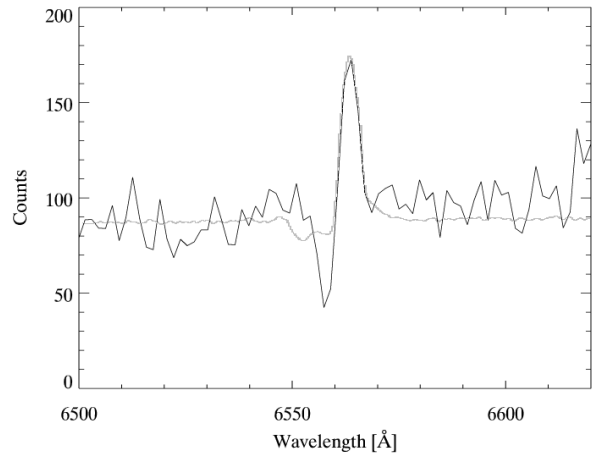}}
        \caption[]{H$\alpha$ spectrum of the nebula. The light grey line shows the spectrum of LMZ\,12 \cite{ReiAsp04} for comparison, scaled to match the emission component.}
        \label{FigSpecNeb}
   \end{figure}  

}

\subsection{The identification of the central source}
\begin{figure*}[t]
   \sidecaption
        \includegraphics[width=12cm]{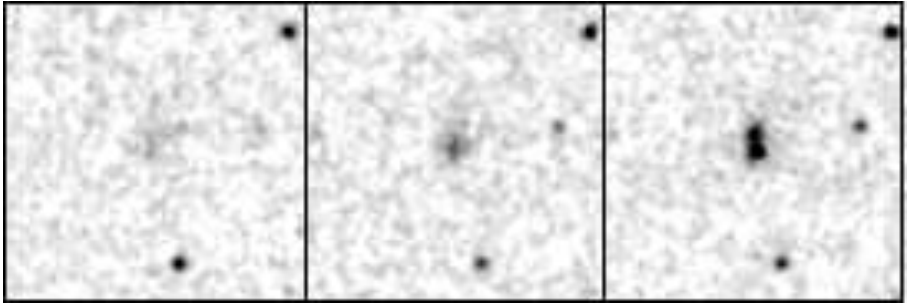}  % \resizebox{\hsize}{!}{\includegraphics[]{2mass_fig.eps}}
        \caption[]{2MASS $J$, $H$, and $Ks$ images (left to right) of the IR source (FOV 1\farcm5$\times$1\farcm5). The intensity range is [$-$1, 15]$\sigma$. North is up and east to the left.}
        \label{Fig3}
   \end{figure*}

The 2MASS images show a very red object located just south of the optical nebula (cf. Fig.\,\ref{Fig2}). It is the reddest object in this subregion of the dark cloud. Since it changes its morphology from 1.24\,$\mu$m to 2.16\,$\mu$m it is likely of non-stellar origin (Fig.\,\ref{Fig3}). In the $J$-band, only faint extended emission is visible. The object becomes elongated in the $H$-band in north-south direction and appears as a double source in {$Ks$} with a position angle of 3\degr{} and a separation of 5\farcs2 of the components. {Another indication for this behaviour is given by the photometric quality flags of the 2MASS point source catalogue which indicate upper limits for both components in $J$, and an upper limit for the northern component in $H$. The northern component almost coincides with the optical brightness peak.} Since the object is the only red source located within the positional error ellipse of IRAS\,04376+5413 it is the near-infrared (NIR) counterpart of the IRAS point source, refered to as L1415-IRS in the following. The IR properties, the association with the optical nebula and the HH flow qualify this source as a young object which is surrounded by an accretion disk. {The NIR magnitudes of L1415-IRS were measured on the 2MASS images using the same size and position of the synthetic aperture as for the optical frames. The magnitudes and IRAS fluxes are compiled in Tab.\,\ref{table:1}. They are not corrected for foreground extinction. 
%There is a marginal trend from the second epoch photometry for a slight brightness decrease.
}

\begin{table}
\caption{Magnitudes/Flux densities of L1415-IRS} % title of Table
\label{table:1} % is used to refer this table in the text
\centering % used for centering table
\begin{tabular}{l l l l} % centered columns (4 columns)
\hline\hline % inserts double horizontal lines
Wavelength& Brightness & Epoch & Comment \\ % table heading
 $\rm [\mu m]$  & [mag]/[Jy] &       &         \\ % table heading
\hline % inserts single horizontal line
0.7 & 15.87$\pm$0.08 & 2006.3094 &  \\ % inserting body of the table
0.7 & 16.06$\pm$0.06 & 2006.6598 &  \\ % CCD_08_2006
0.7 & 16.03$\pm$0.06 & 2006.7473 &  \\ % CCD_09_2006
0.7 & 16.04$\pm$0.06 & 2006.7937 &  \\ % CCD_10_2006
0.9 & 18.37$\pm$0.24 & 1996.9254 &  \\
0.9 & 18.36$\pm$0.70 & 2001.0602 & nondetection \\
0.9 & 14.58$\pm$0.10 & 2006.2519 &  \\
0.9 & 14.61$\pm$0.07 & 2006.3094 &  \\
0.9 & 14.78$\pm$0.06 & 2006.6598 &  \\ % CCD_08_2006
0.9 & 14.74$\pm$0.05 & 2006.7473 &  \\ % CCD_09_2006
0.9 & 14.74$\pm$0.05 & 2006.7937 &  \\ % CCD_10_2006
1.24 & 15.02$\pm$0.08 & 1998.9774 &  \\
1.66 & 13.68$\pm$0.05 & 1998.9774 &  \\
2.16 & 12.39$\pm$0.03 & 1998.9774 &  \\
12  & 0.15$\pm$0.05\,Jy &  1983.5   & mean IRAS epoch \\
25  & 0.49$\pm$0.06\,Jy &  1983.5   & mean IRAS epoch \\
60  & 1.40$\pm$0.30\,Jy &  1983.5   & mean IRAS epoch \\
\hline %inserts single line
\end{tabular}
\end{table}

%______________________________________________________________
\section{Discussion}
\subsection{The distance to LDN\,1415}
Knowledge of the distance to LDN\,1415 is crucial for the interpretation of the results. For this purpose, data of the CO survey of \cite{Dame01} were inspected. Along the line of sight towards LDN\,1415 there are two emission components, a stronger one at $v_{\rm LSR}=-5.2$\,km/s and a weaker one at $v_{\rm LSR}=0$\,km/s. The negative velocity implies a kinematic distance of a few kpc while a velocity close to zero points to a local origin. {However, the kinematic distances of nearby clouds are subject to considerable uncertainties because of the influence of non-neglible peculiar motion.}
%The far distance would imply that the object is rather luminous. In this case it can be expected that is situated in a cluster environment which is not observed. Thus it seems reasonable that the object is at the near distance, and the value for the Taurus-Auriga association of 140\,pc is assumed in the following.
{LDN\,1415 is amidst a complex of dark clouds ranging from LDN\,1387 to LDN\,1439 which stretches about 10 degrees at the border between Camelopardalis and Auriga. Thus, it is reasonable to assume that LDN\,1415 is part of this complex. Another member, LDN\,1407, was included in the molecular line and extinction study of \cite{Snell81} who measured a $^{12}$CO $v_{\rm LSR}$ of 1.5\,km/s and derived a distance of 170\,pc. A margin of $\pm$30\,pc seems to be plausible if the depth of the cloud complex is similar to its extent in the plane of the sky. For another dark cloud of the complex (LDN\,1439, CB\,26), \cite{launh01} adopted a distance of 140\,pc based on a re-examination of the larger-scale velocity structure. An upper limit to the distance of LDN\,1415 can be derived using the stellar population model of the Galaxy of \cite{Robin03}. From the distance dependence of the cumulative number of stars towards the dark cloud and the limiting magnitude, a prediction on the number of foreground stars can be made. The POSS-II N image was used for this purpose. The total area of the three dense contiguous regions lacking any foreground star amounts to $\sim$25$^{\square}$\arcmin{}. According to the model, this surface area implies a maximum distance of 190\,pc for an average value of 0.5 foreground stars, and 250\,pc for 1 foreground star. These values are consistent with the result of \cite{Snell81} which is assumed to hold for LDN\,1415.
}

\subsection{The Herbig-Haro flow}
% line ratios:
% SII[6717+6731]/Halpha=0.403222
% OI[6300]/Halpha=0.171051
% SII[6717]/SII[6731]=1.10645
% NII[6583]/OI[6300]=1.62845
{
The presence of HH\,892 is evidence for recent outflow activity in a well-collimated fashion. The spectrum of HH\,892A resembles that of a high-excitation HHO (\cite{raga96}). From the line ratios conclusions on the excitation conditions can be drawn. The ratio of the [S{\sc\,ii}]$\lambda$6717 and [S{\sc\,ii}]$\lambda$6731 does not depend on the electron temperature and is a good diagnostic of the electron density \cite{czyz86}. It amounts to $n_e=300\pm120\,$cm$^{-3}$. The ratio of the [N{\sc\,ii}]$\lambda$6583 and [O{\sc\,i}]$\lambda$6300 lines traces the ionisation fraction almost independently of pre-shock density and magnetic field strength (\cite{hart94} 1994). The derived value for the ionisation fraction of 0.35$\pm$0.05 implies a total gas density of about 900\,cm$^{-3}$. In order to estimate the shock velocity the ratios [O{\sc\,i}]$\lambda$6300/H$\alpha$, S{\sc\,ii}]$\lambda$,$\lambda$6717$+$6731/H$\alpha$, and [N{\sc\,ii}]$\lambda$6583/[O{\sc\,i}]$\lambda$6300 can be used. Assuming the low-preshock density case of \cite{hart94} (1994) a value of 60$\pm$10\,km/s was derived. Fig.\,\ref{Fig1} shows faint bended H$\alpha$ emission north of HH\,892. It needs to be clarified whether it is scattered off from an outflow wall or due to a precessing jet. The similarity of the radial velocities of both HHOs suggests that the outflow axis is close to the plane of the sky. 
}

{
\subsection{The new optical nebula}
The brightest part of the new nebula has an arcuate morphology. This kind of appearance is often met among YSOs (e.g., \cite{Padg99}, \cite{Zinn99}). Ring-shaped nebulae are archetypical for FU Ori-type objects (FUors). This was first noticed by \cite{Good87} who suggested that they result from light scattering in the lobe of a bipolar structure which is oriented towards the observer. The opposite lobe is obscured by dust in the equatorial waist of the structure. Generally, this appearance can be explained by the interplay of the lobe-opening angle, the observers line of sight, and the dependence of the scattering cross-section of the grains on phase angle. It is obvious that in this simple model a complete ring can only be seen pole on while otherwise, an arcuate nebula is observed. Recently, in the case of V1331 Cyg, a star which is often considered to be a peculiar T-Tauri star, \cite{Quanz06} found a secondary scattering arc closer to the star on archival HST images.

The fact that the optical brightness peak almost coincides with the northern 2MASS source and the presence of fainter optical nebulosity north of the arcuate nebula suggest that the northern outflow lobe is inclined towards us. However, the appearance of the nebula also depends on the location of the source with the respect to the cloud and its shape. An extinction lane seems to run from southwest to northeast across the northern nebulosity (cf. Fig.\,\ref{Fig2}). This is very similar to the case of CB26\,YSO\,1 (\cite{Steck04}). The diffuse H$\alpha$ emission from the northern part of the  nebula (cf. Fig.\,\ref{Fig1}) {traced in the optical spectrum} is presumably scattered light from the accretion shock and the inner disk. 
}

\subsection{L1415-IRS -- a young low-luminosity object}
The NIR morphology of {L1415-IRS} is suggesting a bipolar source, with two outflow lobes seen at 2.16\,$\mu$m (Fig.\,\ref{Fig3}). If the binary appearance in the 2MASS {$Ks$} image is interpreted in this way, an inclination of almost 90\degr{} seems likely. Deviations from this value will cause brightness differences of the lobes because of differing scattering angles. {In the present case, the southern component is brighter by 0.4\,mag at $Ks$ and the photocentres of the $H$- and $J$-band emission are almost coincident with the southern component. This would suggest that the southern lobe is facing towards us. This is in contradication with the conclusion drawn above. While it seems unlikely that the northern 2MASS source is an unrelated object it might be that it is due to a combination of scattered light and shock-excited H$_2$ emission from a jet knot.
}
% However, both the optical nebula and the bright HHO are to the north of the IR source, indicating just the opposite case. 
If the 2MASS images show a young bipolar object indeed, it would be another proof of its proximity. The angular resolution of 2MASS is not sufficient to resolve a YSO at a distance of a few kpc. {The same argument was brought up by \cite{Huard06} for the low-luminosity source L1014-IRS in LDN\,1014. The angular size of the cometary nebula associated with this source in the $Ks$-band is about 10\arcsec{} which matches the extent of L1415-IRS at this wavelength.}  
% {This can be verified by polarimetric observations and provides the chance for a detailed study of the outflow properties, like in the case of the very luminous YSO IRAS\,08159-3543 \cite{NeSt95}.}

\begin{figure}[h]
        \resizebox{\hsize}{!}{\includegraphics[]{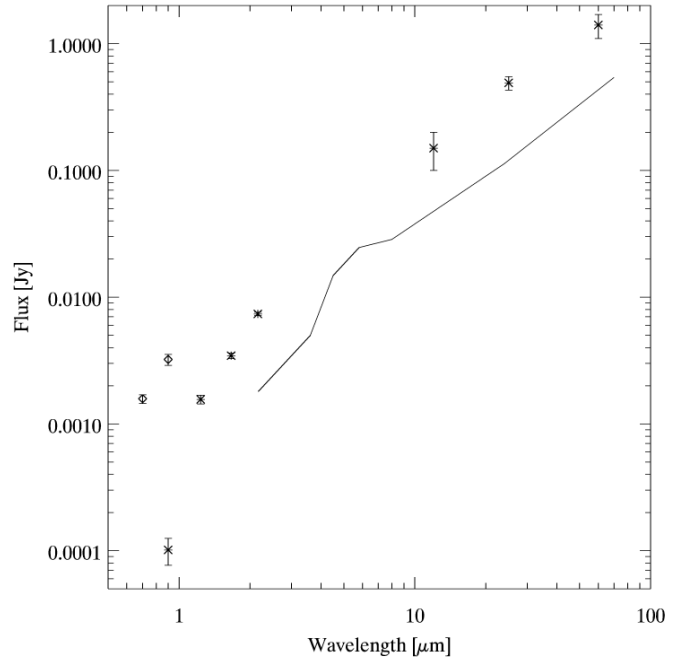}}
        \caption[]{Spectral energy distribution of the embedded object based on the values given in Tab.\,\ref{table:1}. {Asterisks mark the low-state fluxes while diamonds represent refer to the high-state values. The line represents the SED of L1014-IRS (\cite{Young04}), scaled to the distance of LDN\,1415.}}
        \label{Fig4}
   \end{figure}

The spectral energy distribution of L1415-IRS is shown in Fig.\,\ref{Fig4}. {For the conversion of magnitudes to fluxes the relations of \cite{Cohen03} and \cite{Bess90} were used.} The spectral slope indicates a Class I source (\cite{Adams87}). The IRAS 100\,$\mu$m measurement was not included since the scan profile is broader than the PSF, suggesting that the bulk of the emission at this wavelength arises from the dark cloud, presumably due to external heating. The epoch of the 2MASS data is 1998, i.e. in between of the DSS2 and the Kiso observations. Therefore it is reasonable to assume that the 2MASS fluxes are representative for the pre-outburst state. The luminosity {during the inactive state integrated} from 0.9\,$\mu$m to 60\,$\mu$m amounts to {0.13\,$L_{\sun}$ for the assumed distance of 170\,pc. It is comparable to that of the low-luminosity source L1014-IRS (\cite{Young04} 2004), a possible substellar object whose SED is displayed in Fig.\,\ref{Fig4} for comparison. The bolometric luminosity of L1014-IRS amounts to 0.3\,$L_{\sun}$, with the central source contributing 0.09\,$L_{\sun}$ according to the model of \cite{Young04} (2004).  In the comparison, it has to be kept in mind that the luminosity of  for L1014-IRS is based on Spitzer observations while the estimate for L1415-IRS was derived from IRAS data since Spitzer measurements are lacking yet. It might well be that the large IRAS beam is picking up emission which does not stem from the central source.

Although it seems tempting to conclude that L1415-IRS is of similar nature as L1014-IRS, 
% While this would imply that the infrared source might be a substellar object
the uncertainty of the luminosity estimate caused by both the non-isotropic radiation field due to the presence of a disk and the lack of an accurate distance measurement must be kept in mind. The current observational constraints together with the evidence for ongoing accretion do not allow to conclude rigorously that L1415-IRS will eventually become optically visible as a brown dwarf. 
}

% {The youth of the object, the observed brightness variation and the morphology of the source are the key features of L1415-IRS}. The fact that the HHO can be seen in the red DSS2 image which is lacking the nebula indicates that the outflow already penetrated the cloud core at the time this plate was taken. {Thus, in principle photons could have escaped through this channel of reduced optical depth. However, the small luminosity during the low state precluded the detection of the nebula.}

\subsection{FUor- or EXor-like activity of L1415-IRS}
{The presence of the P Cygni profile of the H$\alpha$ line of the nebula provides clear evidence for an} FUor- or EXor-type outburst due to temporarily enhanced accretion.{The brigthening of FUors and EXors results from the increase of the accretion luminosity and the decrease of the extinction since dust grains are blown away by the strong neutral wind \cite{ReiAsp04} }. In the recent past, two similar events were observed. In the year 2000, the $Ks$ brightness of an embedded YSO in the Cygnus OB7 molecular cloud increased by at least 4 magnitudes which, consequently, led to the optical detection of the Braid nebula (\cite{Movs06}).  In early 2004, the outburst of IRAS\,05436-0007, a Class 0 source of 2.7\,$L_{\sun}$ (\cite{Bric06}), raised the visible brightness of its surrounding nebula by about 5 magnitudes, allowing the detection by \cite{McNeil04}. {The H$\alpha$ line of its driving source LMZ 12 \cite{ReiAsp04} is shown for comparison in Fig.\,\ref{FigSpecNeb}. Compared to this source the neutral wind from L1415-IRS has a stronger optical depth but its terminal velocity is smaller. }

In order to explain the FU Orionis phenomenon, several models were brought up. A viscous thermal disk instability has been invoked by \cite{BellLin94}. However, the temperatures 
%(3000$\dots 10^4$\,K)
required for this kind of instability may not be reached in the disk surrounding the cool low-mass object in LDN\,1415. Another explanation is based on the tidal interaction with a companion \cite{BonBast92}, implying periodic outbursts. Recently, numerical simulations of protostellar disk evolution showed that gravitational torques may lead to the swallow of protoplanetary embryos by the protostar (\cite{VorBas06}), causing repetetive bursts of accretion. This process requires continuous infall from the envelope to keep the disk gravitationally instable. For what concerns L1415-IRS, this condition seems to be met since it is still embedded and surrounded by a halo of matter. Recent studies show that the disks of brown dwarfs are quite similar to those of low- and intermediate-mass stars (e.g., \cite{Natta04}, \cite{Apai05}). The observation of an FUor-type event from a low-luminosity YSO adds another piece of evidence to this finding. 

At present we are witnessing an outburst of L1415-IRS -- {a cry from the cradle of a low-luminosity object}.
{
The photometry in 2006 listed in Tab.\,\ref{table:1} which covers almost 200 days shows only a marginal trend for a brightness decrease and no evidence for strong fluctuations. For comparison, the high-state of EX Lupi in 1955-1956 lasted about 250 days (\cite{Herbig77}).
}
Future brightness monitoring and spectroscopy will prove whether L1415-IRS belongs to the class of FUors or shares more similarities with EXors. Furthermore, it will allow to investigate the interaction of the disk wind with the infalling envelope \cite{Clark05}.

So far, there was no object among the sparse sample of FUors and EXors which is as less luminous as L1415-IRS presented here. {The average bolometric luminosity of the FUors listed by \cite{HaKa96} amounts to $\sim$250\,L$_{\sun}$. }
% Of similar interest are molecular line and dust continuum observations. 
Sensitive submillimetre mapping may reveal if not only the SED but also the outflow properties of L1415-IRS are similar to those of L1014-IRS (\cite{Bourke06}). Dust continuum observations are needed to constrain the bolometric luminosity and models of the disk/envelope. Finally, NIR spectroscopy is of particular importance since it may reveal the presence of the 2.3\,$\mu$m CO band-head absorption which is a strong spectroscopic signature for FUors \cite{SanWein01}.
%will yield an estimate of the accretion rate $\dot M_{acc}$. % In this context, it would be of interest whether L1415-IRS fits the $\dot M_{acc} - M_{\ast}$ relation of \cite{Natta06} for PMS stars. }  
%______________________________________________________________

\section{Conclusions}

   \begin{enumerate}
      \item By means of H$\alpha$ and [S{\sc\,ii}] imaging as well as long-slit spectroscopy the HH nature of an emission-line dominated object in LDN\,1415, first seen on a red DSS2 plate, could be verified. {This object and a fainter one constitute the HH\,892 flow which is a sign of recent accretion and outflow activity of an embedded young object.}
      \item A new compact optical nebula {in LDN\,1415 was found. The nebula is associated with a Class I source, L1415-IRS}, which shows a bipolar appearance at near-infrared wavelengths. For the likely distance of 170\,pc, its luminosity amounts to 0.13\,$L_{\sun}$.
      \item The brightness increase of the nebula by about 3.8\,mag in recent years is due to an FUor- or EXor-type event. {This is confirmed by the presence of a P Cygni profile of the H$\alpha$ line}. {The occurence of this kind of activity in a circumstellar disk surrounding a very low-luminosity object may provide a clue to the general understanding of this phenomenon.}
   \end{enumerate}

\begin{acknowledgements}
{Observational support by F. Ludwig and U. Laux is gratefully acknowledged. Comments from the anonymous referee, D. Froebrich, and H. Linz helped to improve the paper.} This publication makes use of data products from the Two Micron All Sky Survey, which is a joint project of the University of Massachusetts and the Infrared Processing and Analysis Center/California Institute of Technology, funded by the National Aeronautics and Space Administration and the National Science Foundation. Based in part on data collected at Kiso observatory (University of Tokyo) and obtained from the SMOKA science archive at Astronomical Data Analysis Center, which are operated by the National Astronomical Observatory of Japan. This research has made use of the NASA/IPAC Infrared Science Archive, which is operated by the Jet Propulsion Laboratory, California Institute of Technology, under contract with the National Aeronautics and Space Administration.
\end{acknowledgements}

\rm

\end{document}